\documentclass[twocolumn,aps,floatfix]{revtex4}
\usepackage{amssymb}
\usepackage{graphicx}

\begin{document}

\title{Dynamics of optically generated vortices in a one--component
       ultracold fermionic gases}

\author{Tomasz Karpiuk,$^1$ Miros{\l}aw Brewczyk,$^1$ and 
        Kazimierz Rz\c a\.zewski$\,^2$}                          

\affiliation{\mbox{$^1$Uniwersytet w Bia{\l}ymstoku, ul. Lipowa 41,
                        15-424 Bia{\l}ystok, Poland}  \\
\mbox{$^2$Centrum Fizyki Teoretycznej PAN and College of Science,
           Al. Lotnik\'ow 32/46, 02-668 Warsaw, Poland}  }             

\date{\today}

\begin{abstract}
We show that the phase imprinting method is capable of generating
vortices in a one--component gas of neutral fermionic atoms at zero 
and finite temperatures. We find qualitative differences in dynamics
of vortices in comparison with the case of Bose--Einstein condensate.
The results of the imprinting strongly depend on the geometry of the 
trap, e.g., in asymmetric traps no single vortex state exists. These 
observations could be considered as a signature of Cooper--pair based 
superfluidity in a Fermi gas.

PACS number(s): 05.30.Fk, 03.75.Fi \newline
\end{abstract}

\maketitle

One of the spectacular properties of a macroscopic bosonic--type quantum
system is a superfluidity, originally observed in a liquid helium II.
Although superfluidity is a complex phenomenon, it is certainly intimately
related to the existence of quantized vortices. The recent experimental
realization of Bose--Einstein condensation in confined alkalai--metal gases
\cite{BEC} allowed to study this connection in details. Quantized vortices 
(in a form of small arrays as well as completely ordered Abrikosov lattices) 
have been already observed in many laboratories \cite{vortex}. Interferometric 
detection method showed directly $2\pi$--phase winding associated with the
presence of a vortex in the condensate \cite{Dalibard}. Of course, this is a
manifestation of the macroscopic wave function.

However, the degenerate Fermi gas has no macroscopic wave function. It is 
rather described (at zero temperature) in terms of the many--body wave
function, built of single--particle orbitals, whose evolution is governed 
by the Schr\"odinger equation. Hence, the appearance of quantized vortices 
in a fermionic system demonstrates the BCS--type transition from normal to 
superfluid phase. Recently several groups have undertaken the effort to 
achieve quantum degeneracy in a dilute Fermi gas \cite{Jin,Hulet,Salomon}. 
In JILA experiment \cite{Jin,Jin1} $^{40}$K atoms were trapped in two 
hyperfine states and cooled evaporatively by collisions between atoms in a 
different spin states whereas in Refs. \cite{Hulet,Salomon} a mixture of 
bosonic ($^7$Li) and fermionic ($^6$Li) atoms was used to perform the 
sympathetic cooling of fermions. 

In this Letter we investigate the dynamics of optically generated vortices
in a Fermi gas in a normal phase and compare it with the properties
of vortices in the Bose--Einstein condensate. A good approximation to
the spin--polarized Fermi gas at low temperatures is that of non--interacting
particles (the s--wave scattering is absent for spin--polarized fermions). 
We then assume that at zero temperature the system is described by 
the Slater determinant with the lowest orbitals being occupied. The
details of the technique of phase imprinting are thoroughly discussed
in Refs. \cite{Dobrek,Andrelczyk} for bosons and repeated in Refs.
\cite{Tomek1,Tomek2} for fermions. It is worth stressing that this method
was already experimentally realized for the generation of solitons 
in Bose--Einstein condensate \cite{Hannover,NIST}.

We start our considerations with a two--dimensional symmetric trap and
imprint the phase at the center of the trap. Each atom acquires a phase
given by $k \phi$, where $k$ is an integer number and $\phi$ is the 
azimuthal angle around the origin (each wave function is multiplied by 
the factor $\exp{(i k \phi)}$). The following free evolution of the 
single--particle orbital can be calculated by using the propagator 
technique
\begin{equation}
\varphi_j (x,y,t) = \int \int K(x,y,x^{\,\prime},y^{\,\prime},t) \,
\varphi_j (x^{\,\prime},y^{\,\prime},0) \, dx^{\,\prime}dy^{\,\prime} \, ,
\label{evolution}
\end{equation}
where $K(x,y,x^{\,\prime},y^{\,\prime},t)$ is the propagator function for
two--dimensional symmetric harmonic oscillator \cite{Feynman}
\begin{widetext}
\begin{eqnarray*}
K(x,y,x^{\,\prime},y^{\,\prime},t)=\frac{{\rm M} \omega}{2\pi \hbar i 
\sin{\omega t}} \exp{\left\{\frac{i {\rm M} \omega}{2 \hbar \sin{\omega t}} 
\left[ (x^2 + y^2 + x^{\,\prime\,2} + y^{\,\prime\,2})
\cos{\omega t} - 2\, x x^{\,\prime} -2\, y y^{\,\prime} \right] \right\} } \;.
\end{eqnarray*}
\end{widetext}
In this case it is convenient to represent the single--particle orbitals in 
polar coordinates (hereafter, $\sqrt{\hbar/({\rm M}\, \omega)}$ and $1/\omega$ 
are used as units of length and time, respectively):
$\varphi_{n m} (r,\phi) = A_{nm} \, r^{|m|} \: L_n^{|m|} (r^2) \:
e^{-r^2 /2} \, e^{i m \phi}$  and $A_{nm} = (n! / [\pi (n+|m|)!])^{1/2}$
and the integration according to the prescription (\ref{evolution}) can
be performed first over the azimuthal angle then over the radial distance.

The first stage requires the use of the formula \cite{MathFun}
\begin{equation}
\int \limits_{0}^{\:2\pi} e^{-i z \cos{\Theta}}\, e^{i l \Theta}\, d\Theta = 
(-i)^l\, 2 \pi\, J_l(z)  \;,
\label{formula1}
\end{equation}
where $J_l(z)$ is the Bessel function of integer order and $z$ any complex 
number. In our case $z$ is real and equals $r r^{\, \prime} / \sin{t}$.
In the second stage one has to calculate
\begin{equation}
\int \limits_{0}^{\:\infty} r^{\, \prime \, (1+|m|)}\:
e^{- a^2 r^{\,\prime\,2}} J_{m+k}(b\, r^{\, \prime})\,
L_n^{|m|}\, (r^{\, \prime\,2})\;  d r^{\, \prime}  \:,
\label{calka}
\end{equation}
where $a^2=1/2-i\cos{t}/(2 \sin{t})$ and $b=r/\sin{t}$. Since the generalized 
Laguerre function $L_n^{|m|}$ is a polynomial, the expression (\ref{calka}) 
can be expanded and integrated term by term with the help of appropriate formula 
for integrals of Bessel functions \cite{MathFun}
\begin{eqnarray}
\int \limits_{0}^{\:\infty} r^{\, \prime \, (\mu-1)}
e^{- a^2 r^{\,\prime\,2}} J_{\nu}(b\, r^{\, \prime})\,
d r^{\, \prime} =
\frac{\Gamma(\frac{1}{2}\nu+\frac{1}{2}\mu) (\frac{1}{2}\frac{b}{a})^\nu}
{2 a^\mu \Gamma(\nu+1)}  \nonumber  \\  \times \,
M\left( \frac{1}{2}\nu+\frac{1}{2}\mu,\nu+1,-\frac{b^2}{4 a^2} \right)
\phantom{aaaaaaaaaa}
\label{formula2}
\end{eqnarray}
valid for ${\rm Re}(a^2)>0$ and ${\rm Re}(\mu+\nu)>0$. $M(\alpha,\gamma,z)$ 
is a confluent hypergeometric function.

\begin{figure}[htb]
\resizebox{2.7in}{2.2in}
{\includegraphics{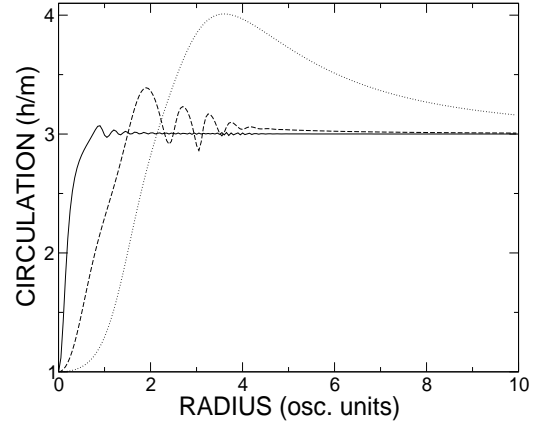}}
\caption{Circulation along the circle as a function of its radius for
various moments: $0.05$ (solid line), $0.26$ (dashed line), and $1.57$ 
(dotted line) in units of $1/\omega$. The system and phase imprinting 
parameters are as follows: $N=6$ and $k=3$.}
\label{circulation}
\end{figure}

The result is the product of $b^{m+k}$ and the combination of the confluent 
hypergeometric functions; each of the argument equal to $- b^2 / (4 a^2)$ 
\begin{eqnarray*}
\varphi(r,\phi,t) = -\frac{A_{nm} (-i)^{m+k-1}}{\sin{t}}\, 
e^{i \frac{\cos{t}}{2 \sin{t}} r^2} e^{i (m+k) \phi}
\phantom{aaaa}   \nonumber  \\
\times \frac{b^{m+k}}{\Gamma(1+m+k)\: 2^{1+m+k}} 
\phantom{aaaaaaaaaaaaaaaaaaaaiii}   \nonumber  \\
\times \sum_{l=0}^{n} \frac{(-1)^l}{l!}  
\left( \begin{array}{c}
          n+|m| \\ n-l
       \end{array}      \right) 
\frac{\Gamma(1+l+\frac{m+k+|m|}{2})}{a^{2+2l+m+k+|m|}}  
\phantom{aaaai} & & \nonumber  \\
\times\, M\left( 1+l+\frac{m+k+|m|}{2},1+m+k,-\frac{b^2}{4 a^2} \right)  \;.
\phantom{ai}
\end{eqnarray*}
Hence, the value of the wave function (single--particle orbital) at the 
origin is fully determined by the factor $b^{m+k}$ and equals zero provided 
the phase imprinting is strong enough ($k$ is larger than the maximum absolute 
value of magnetic number $m$ of occupied orbitals). In such a case the 
diagonal part of the one--particle density matrix is zero at the origin all 
the time. On the other hand, the part of the phase of particular orbital 
dependent on the azimuthal angle is given by $(m+k) \phi$ and the circulation
along the circle centered at the vortex line (point) is \cite{Tomek1}  
\begin{eqnarray*}
\Gamma_C(r,t) = \frac{h}{{\rm M}}\: \frac{\sum_{j} |\varphi_{j}(r,t)|^2 (m_j +k)}
{\sum_{j} |\varphi_{j}(r,t)|^2}  \;.
\end{eqnarray*}
In Fig. \ref{circulation} we plotted the circulation for various paths 
(circles centered at the vortex point) at different times for a system 
of $N=6$ fermions after writing the phase $3 \phi$. It is clear that the 
strong enough phase imprinting leads to the excitation of the Fermi gas 
with non--zero (but not quantized) circulation around the point where the 
density vanishes. We call this state the vortex. 

For weaker phase imprinting (e.g., for $k=2$ and $N=6$) the response of 
the system becomes more complex. The density just does not go to zero at the 
center. In some cases the density at the origin oscillates between zero
and local maximum and the circulation goes to zero while the radius of the 
circle decreases (vortex--antivortex oscillations). In others, the dip
in the density is strongly diminished, however the vortex reappears at
multiple integers of the trap period.

It turns out that we can follow analytically the evolution of the single--particle
orbitals also in the case when the absorption plate (and the laser beam) is
shifted off the center of the trap. For that, the single--particle orbitals are
multiplied by the factor
\begin{eqnarray*}
\left[    \frac{(x-x_0) + i (y-y_0)}
               {\sqrt{(x-x_0)^2 + (y-y_0)^2}}    \right]^k    \;,
\end{eqnarray*}
where $(x_0,y_0)$ are coordinates of the laser beam with respect to the
center of the trap. This time it is better to represent the atomic wave 
functions in Cartesian coordinates: $\varphi_{n_1 n_2}(x,y)=A_{n_1} A_{n_2}
H_{n_1}(x) H_{n_2}(y) e^{-(x^2+y^2)/2}$, where $A_n=(1/(\pi^{1/2} 2^n n!))^{1/2}$
and $H_n(x)$ is the n--th Hermite polynomial.

To find the evolution of single--particle orbitals, governed by Eq. (\ref{evolution}), 
it is convenient to shift the origin of Cartesian coordinates 
($x^{\,\prime},y^{\,\prime}$) by the vector $(x_0,y_0)$ and then introduce 
the polar coordinates. The integration over the azimuthal angle requires the 
use of generating function technique and the following extension of formula 
(\ref{formula1})
\begin{eqnarray*}
\int \limits_{0}^{\:2\pi} e^{i (z \cos{\phi} + w \sin{\phi})} \,
e^{i k \phi}\, d\phi = \frac{2\pi i^k (z+iw)^k}{(\sqrt{z^2+w^2})^k}
J_k(\sqrt{z^2+w^2})
\end{eqnarray*}
with $z$ and $w$ being complex numbers. The result $({\rm I}_{n_1n_2}(r^{\,\prime}))$
for any state $(n_1,n_2)$ is obtained by expanding the left--hand side of the
equation
\begin{eqnarray}
e^{-(s_1^2+s_2^2)}  e^{2 (s_1 x_0 + s_2 y_0)}
\frac{2\pi i^k (z+iw)^k}{(\sqrt{z^2+w^2})^k} 
J_k(r^{\,\prime} \sqrt{z^2+w^2}) =  \nonumber  \\
\sum_{n_1=0}^{\infty} \sum_{n_2=0}^{\infty} \frac{1}{n_1!} \frac{1}{n_2!} \,
{\rm I}_{n_1n_2}(r^{\,\prime})\,  s_1^{n_1} s_2^{n_2}  \;,
\phantom{aaaaaaaaaa}
\label{genfun}
\end{eqnarray}
where
\begin{eqnarray*}
z & = & \frac{1}{\sin{t}} ( x_0 e^{it} -x) -i 2 s_1  \nonumber \\
w & = & \frac{1}{\sin{t}} ( y_0 e^{it} -y) -i 2 s_2  \nonumber \\
\end{eqnarray*}
in a Taylor series with respect to $(s_1,s_2)$ around $(0,0)$ point.

\begin{figure}[htb]
\resizebox{2.9in}{2.9in}
{\includegraphics{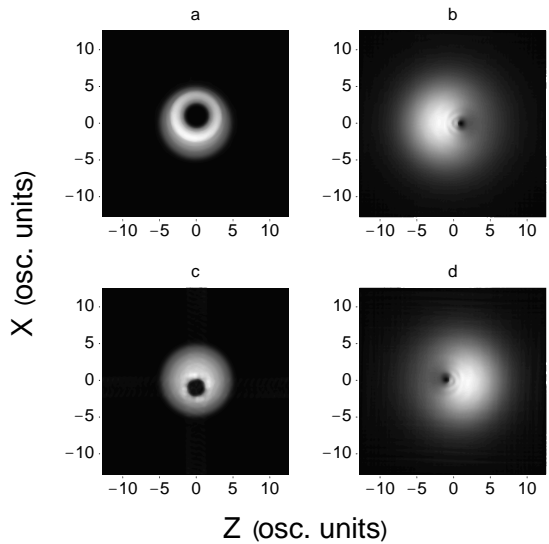}}
\caption{Evolution of the density distribution of $N=105$ atoms in 
two--dimensional symmetric harmonic trap after imprinting a phase of 
($14\, \phi$)\, $1.5$ osc. units off the center. The successive 
frames correspond approximately to the moments: (a) 0, (b) 1/4, 
(c) 1/2, and (d) 3/4 of trap period. The vortex rotates around the
center of the trap changing periodically the size of its core.}
\label{vortex4}
\end{figure}

Since the only dependence on the distance $r^{\,\prime}$ appears through 
the $J$ Bessel function it is convenient to perform the integration over 
the radial distance first. Again, by using the formula (\ref{formula2}) one 
gets (up to time--dependent constant)
\begin{equation}
e^{-(s_1^2+s_2^2)}  e^{2 (s_1 x_0 + s_2 y_0)}\, (z+iw)^k
\label{genfun1}
\end{equation}
for the left--hand side of (\ref{genfun}). The expression (\ref{genfun1})
implies that, assuming the phase imprinting is strong enough 
$(k > {\rm max}(n_1,n_2))$,
each atomic wave function includes the factor $(z+i w)$ in an integer power 
higher than or equals $1$, taken at the point $(s_1=0,s_2=0)$. It is easy to 
check that
\begin{eqnarray}
(z+iw)_{s_1=0,s_2=0} = (x_0 \cos{t} - y_0 \sin{t} - x) + 
\phantom{aa}  \nonumber  \\
i\,(x_0 \sin{t} + y_0 \cos{t} - y)  \;.
\phantom{aaii}
\label{zero}
\end{eqnarray}
It means that for $k > {\rm max}(n_1,n_2)$ all orbitals have zero at the 
point \mbox{$(z+iw)_{s_1=0,s_2=0} = 0$}. So, this zero is also present in 
the diagonal part of the one--particle density matrix. Moreover, it is clear
from (\ref{zero}) that the zero rotates around the center of the trap at 
the radius equal to $\sqrt{x_0^2+y_0^2}$ with constant frequency which is 
the trap frequency. For $n_1=n_2=0$ the time--dependent wave function 
is given by
\begin{eqnarray*}
\varphi(x,y,t) = - \frac{\Gamma(1+\frac{k}{2})}{\Gamma(1+k)}
\frac{A_{0} A_{0} i^{k-1}}{2^{1+k} a^{2+k} \sin{t}}\,
e^{i \frac{\cos{t}}{2 \sin{t}} (r^2+r_0^2) - \frac{1}{2} r_0^2}
\phantom{aaaaii} \nonumber  \\
\times\, e^{i \frac{1}{\sin{t}}  (x x_0+y y_0)}
(z_1 + i w_1)^k \,
M\left( 1+\frac{k}{2},1+k,-\frac{b^2}{4 a^2} \right)  \;,
\phantom{aaaaii}
\end{eqnarray*}
where $z_1=(x_0 \exp{(i t)} -x)/\sin{t}$, $w_1=(y_0 \exp{(i t)} -y)/\sin{t}$,
and $b^2 = z_1^2 + w_1^2$. Evolution (obtained numerically by solving the 
many--body Schr\"odinger equation \cite{Tomek1,Tomek2}) of the density 
distribution for $N=105$ atoms is shown in Fig. \ref{vortex4}.

\begin{figure}[bht]
\resizebox{2.9in}{2.4in}
{\includegraphics{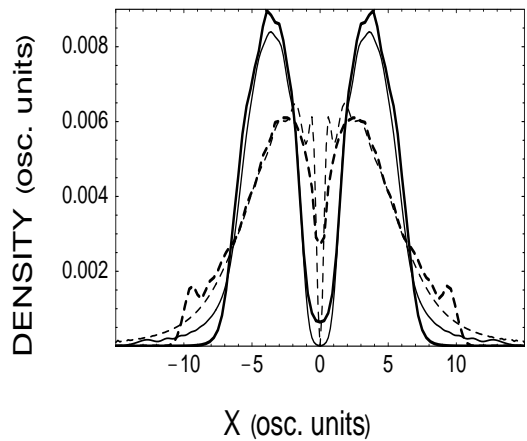}}
\caption{Density distribution (cuts along 'x' axis) of $N=105$ (average 
number) atoms at temperatures zero and $0.1 T_F$ after writing the phase 
$14 \phi$. The curves correspond to moments $\:0.75$ (thick and thin solid 
lines for $0.1 T_F$ and zero temperature, respectively) and $1.5$ (thick and 
thin dashed lines  for $0.1 T_F$ and zero temperature, respectively) in units 
of $1/\omega$ and are the results of averaging procedure over $1000$ 
configurations. The thick dashed curve shows the moment of the lowest contrast 
(which is about $50 \%$).}
\label{temp}
\end{figure}

We consider now the generation and dynamics of vortices at finite
temperatures. To this end, we employ the Monte Carlo algorithm and
generate a number of many--body configurations \cite{Tomek2}
according to the Fermi--Dirac statistics. At finite temperatures
the single--particle states with the energy above the Fermi level
are populated. It may happen that for a given phase imprint $k$,
even strong enough to generate a vortex at zero temperature, there 
are configurations with such single--orbitals that $m+k=0$. For these 
configurations we will see again complex structures like vortex--antivortex 
oscillations in the center of the trap or circulating around it, 
depending on the way the phase imprinting was done. Averaging procedure, 
according to the grand canonical ensemble rules just shows the vortex 
with the lower contrast (see Fig. \ref{temp}). However, one might argue
that in a particular experimental realization only one configuration
is involved and hence vortices, reappearing vortices or vortex-antivortex
oscillations should be observed. In any case, non of these structures
is dissipatively circulating out of the system as predicted for the
vortices in the Bose--Einstein condensate at finite temperatures
\cite{Fetter}.

\begin{figure}[thb]
\resizebox{2.9in}{2.9in}
{\includegraphics{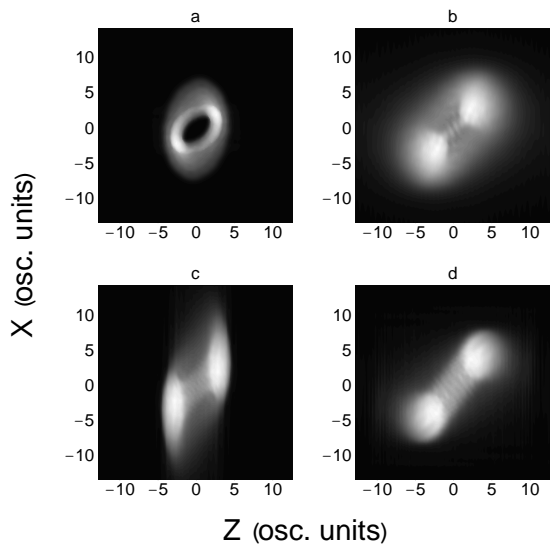}}
\caption{Evolution of the density distribution of $N=105$ atoms in 
two--dimensional asymmetric harmonic trap $(30{\rm Hz} \times 50{\rm Hz})$
after imprinting a phase of ($14\, \phi$)\, at the center. The successive 
frames correspond approximately to the moments: (a) 0, (b) 1/4, (c) 1/2, 
and (d) 3/4 of trap period. The vortex reappears at integer multiples of 
trap period.}
\label{novortex}
\end{figure}

Finally, we have done some numerical calculations for asymmetric traps at
zero temperature. The results are presented in Fig. \ref{novortex}. Again, 
striking difference in comparison with the Bose--Einstein condensate case 
is found. No vortex is generated, presumably because there is no common
zero for orbitals after phase imprinting in the case of asymmetric trap.

In conclusion, we have shown that the dynamics of vortices generated 
in a cold gas of neutral fermionic atoms significantly differs from the 
case when they are generated in the Bose--Einstein condensate. At zero
temperature and symmetric harmonic trap, the vortex rotates around the
center of the trap. However, for fermions it periodically changes the
size of its core. For bosons, the size of the vortex core remains at
the order of the healing length. Further differences appear at finite
temperatures. Contrary to dissipative escape of vortex from the trap
predicted for the Bose--Einstein condensate \cite{Fetter}, we find the stable
rotational motion of the fermionic vortex. Finally, the case of asymmetric 
trap shows new density patterns present in a Fermi gas and not observed
for bosons. All these findings could be used as a test of BCS--type phase
transition in a Fermi gas. \\

\acknowledgments
K.R. acknowledges support of the Subsidy by Foundation for Polish Science.

\end{document}